\documentclass[10pt]{article}
\usepackage{amsmath,amssymb}
\numberwithin{equation}{section}
\usepackage{hyperref}
\usepackage{units}
\usepackage{color}
\usepackage[T1]{fontenc}
\usepackage[utf8]{inputenc}
\usepackage{authblk}
\usepackage{bm}
\usepackage{graphicx}
\usepackage[toc,page]{appendix}

\title{An alternative derivation of the Fern\'andez-Castro analytic approximate expression for the eigenvalues of the bounded quartic oscillator
\thanks{%
PACS:03.65.Ge} 
}
\author[]{Kunle Adegoke\thanks{Corresponding author: adegoke00@gmail.com}}
\author[]{Adenike Olatinwo}
\author[]{Gbenga Olunloyo}

\affil{Department of Physics and Engineering Physics, \mbox{Obafemi Awolowo University}, Ile-Ife, Nigeria}

\begin{document}

\date{}

\maketitle

\begin{abstract}
\noindent In this note we show that the standard \mbox{Rayleigh-Schr\"odinger} (RS) perturbation method gives the same result as the hypervirial pertubative method (HPM), for an approximate analytic expression for the energy eigenvalues of the bounded quartic oscillator. This connection between the HPM and the RS method went unnoticed for a long time, apparently because it was not obvious that the resulting polygamma sums to be evaluated in the RS method could, in fact, be expressed in closed form.
\end{abstract}

\tableofcontents

\section{Introduction}
Interest in the {\em bounded} quartic oscillator started with the pioneering work of Barakat and Rosner~\cite{barakat} who employed a power series method to obtain numerical values of the eigenvalues through an iteration scheme. Researchers have since continued to investigate the bounded quartic oscillator and related systems, using various techniques (see references~\cite{fernandez,navarro,chauduri,alhendi} and the references in them).

\bigskip

The bounded quartic oscillator is described by the Hamiltonian
\begin{equation}
H=-\frac{\hbar^2}{2m}\frac {\rm d^2}{{\rm d}x^2}+\lambda x^4,\quad -a\le x\le a\,,
\end{equation}
where $m$ is the mass of the oscillator and $\lambda>0$ is the coupling constant. The Hamiltonian $H$ lives in a Hilbert space $\mathcal H$ with inner product between any two \mbox{real-valued} functions $f(x)$ and $g(x)$ in $\mathcal H$ defined by $\left(f(x),g(x)\right)=\int_{-a}^a{f(x)g(x)\,{\rm d}x}$, where the functions $f(x)$ and $g(x)$ and indeed all vectors of $\mathcal H$ are required to vanish at the boundary~$x=\pm a$.

\bigskip

About three and a half decades ago, using their hypervirial pertubative method (HPM), Fern\'andez and Castro~\cite{fernandez} derived the following expression (their equation~(22) in our notation) for the eigenvalues of the bounded quartic oscillator:
\begin{equation}\label{equ.dzka6e0}
\begin{split}
E_r  &\approx \frac{{\pi^2 \hbar^2 }}{{8ma^2 }}(r + 1)^2  + \frac{{\lambda a^4 }}{5}\left[ {1 - \frac{{20}}{{\pi^2 (r + 1)^2 }} + \frac{{120}}{{\pi^4 (r + 1)^4 }}} \right]\\
&\qquad + \frac{{32m\lambda ^2 a^{10} }}{{\hbar ^2 }}\left[ {\frac{1}{{225\pi^2 (r + 1)^2 }} - \frac{{37}}{{35\pi^4 (r + 1)^4 }}} \right.\\
&\qquad\qquad\qquad\qquad\left.{  + \frac{{314}}{{5\pi^6 (r + 1)^6 }}- \frac{{1404}}{{\pi^8 (r + 1)^8 }} + \frac{{8712}}{{\pi^{10} (r + 1)^{10} }}}\right]\,,
\end{split}
\end{equation}
for quantum numbers $r=0,1,2,\ldots$

\bigskip

In this paper we show that the standard \mbox{Rayleigh-Schr\"odinger} (RS) perturbation theory with $\lambda$ as the perturbation parameter gives the same result for $E_r$ as given in~\eqref{equ.dzka6e0}. As a matter of fact we stumbled upon the work of Fern\'andez and Castro only after we had obtained our result for $E_r$. The Computer Algebra System Waterloo Maple came to our aid in simplifying the resulting perturbation sums and finding their closed form. 

\section{Basis functions and the matrix elements of~$H$}
Since $H(-x)=H(x)$, the eigenstates of $H(x)$ have definite parity. For $r=0,1,2,\ldots$, the complete orthonormal functions $\{\varphi_r(x)\}$, where,
\[
\begin{split}
\varphi _{2r} (x) = \sqrt {\frac{1}{a}} \cos \left( {\frac{{(2r + 1)\pi x}}{{2a}}} \right),&\quad\varphi _{2r + 1} (x) = \sqrt {\frac{1}{a}} \sin \left( {\frac{{(r + 1)\pi x}}{a}} \right)\,,
\end{split}
\]
constitute a suitable set of basis functions in $\mathcal H$ for a matrix representation of the bounded quartic oscillator Hamiltonian $H$, since they also satisfy the boundary conditions $\varphi_r(\pm a)=0$.

\bigskip

The identities
\[
\begin{split}
\left({\cos \alpha x,\cos \beta x}\right)  = (1 - \delta _{\alpha \beta } )\left( {\frac{{\sin ((\alpha  - \beta )a)}}{{\alpha  - \beta  + \delta _{\alpha \beta } }} + \frac{{\sin ((\alpha  + \beta )a)}}{{\alpha  + \beta }}} \right)&\\
+ \delta _{\alpha \beta } \left( {a + \frac{{\sin (2\alpha a)}}{{2\alpha }}} \right)\qquad&
\end{split}
\]
and
\[
\begin{split}
\left({\sin \alpha x,\sin \beta x}\right)  = (1 - \delta _{\alpha \beta } )\left( {\frac{{\sin ((\alpha  - \beta )a)}}{{\alpha  - \beta  + \delta _{\alpha \beta } }} -\frac{{\sin ((\alpha  + \beta )a)}}{{\alpha  + \beta }}} \right)&\\
+ \delta _{\alpha \beta } \left( {a - \frac{{\sin (2\alpha a)}}{{2\alpha }}} \right)\qquad&\,,
\end{split}
\]
for $\alpha,\beta>0$ and the repeated application of Leibnitz rule for differentiating an integral allow to calculate the matrix elements of $H$ as
\begin{equation}\label{equ.jevxxpj}
\begin{split}
H_{rs}  &= \delta _{rs} \left[ {\frac{{\pi ^2 \hbar ^2 }}{{8ma^2 }}(r + 1)^2  + \frac{{\lambda a^4 }}{5}\left( {1 - \frac{{20}}{{\pi ^2 (r + 1)^2 }} + \frac{{120}}{{\pi ^4 (r + 1)^4 }}} \right)} \right]\\
&\quad + (1 - \delta _{rs} )\left[ {( - 1)^{(r + s)/2} \frac{{16\lambda a^4 }}{{\pi ^2 }}\left( {\frac{1}{{(r - s + \delta _{rs} )^2 }} - \frac{1}{{(r + s + 2)^2 }}} \right)} \right.\\
&\qquad\qquad\qquad\left. { - ( - 1)^{(r + s)/2} \frac{{384\lambda a^4 }}{{\pi ^4 }}\left( {\frac{1}{{(r - s + \delta _{rs} )^4 }} - \frac{1}{{(r + s + 2)^4 }}} \right)} \right]\,,
\end{split}
\end{equation}
provided that $r$ and $s$ have the same parity, and $H_{rs}=0$ otherwise.

\bigskip

The matrix elements $H_{rs}$ facilitate the direct diagonalization of the bounded quartic oscillator. The energy eigenvalues can be made arbitrarily accurate by increasing the dimension of the Hamiltonian matrix used; the eigenvalues obtained can therefore be considered exact. We are, however, not concerned here with exact diagonalization but we need the matrix elements $H_{rs}$ for our perturbation calculations.
\section{RS derivation of the approximate analytic expression for the energy eigenvalues}
For $\lambda$ sufficiently small (see~\cite{navarro} for a rigorous discussion of the convergence criteria), the oscillator potential $V(x)=\lambda x^4$ may be treated as a perturbation of the unperturbed Hamiltonian $T(x)=-\hbar^2/2md^2/dx^2$ (the free particle in a box Hamiltonian). 

\bigskip

In the standard \mbox{Rayleigh-Schr\"odinger} perturbation theory for \mbox{non-degenerate} states, the approximate energy eigenvalues of $H$, to second order in $\lambda$, are to be calculated from $E_r\approx E_r^{(0)}+E_r^{(1)}+E_r^{(2)}$. We have immediately that
\begin{equation}\label{equ.zdjzs3r}
E_r^{(0)}  = T_{rr}  = \frac{{\pi ^2 \hbar ^2 }}{{8ma^2 }}(r + 1)^2=\varepsilon(r+1)^2
\end{equation}
and
\begin{equation}\label{equ.g606140}
E_r^{(1)}  = V_{rr}  = \frac{{\lambda a^4 }}{5}\left( {1 - \frac{{20}}{{\pi ^2 (r + 1)^2 }} + \frac{{120}}{{\pi ^4 (r + 1)^4 }}} \right)\,,
\end{equation}
where $ \varepsilon=\pi ^2 \hbar ^2 /{8ma^2 }\,.$

\bigskip

The second order correction to the energy of the bounded quartic oscillator, $ E_r^{(2)}$, is given by
\begin{equation}\label{equ.kzjql8y}
E_r^{(2)}  = \sum_{\scriptstyle s = 0 \hfill \atop 
  \scriptstyle s \ne r \hfill}^\infty {\frac{{V_{rs} V_{sr} }}{{\varepsilon _{rs} }}}  = \sum_{s = 0}^{r - 1} {\frac{{V_{rs} V_{sr} }}{{\varepsilon _{rs} }}}  + \sum_{s = r + 1}^\infty {\frac{{V_{rs} V_{sr} }}{{\varepsilon _{rs} }}}\,,
\end{equation}
where
\begin{equation}\label{equ.izptbkw}
\varepsilon _{rs}  = \varepsilon _r  - \varepsilon _s  = \varepsilon (r + s + 2)(r - s)\,,
\end{equation}
so that
\begin{equation}\label{equ.cxdserf}
\frac{1}{{\varepsilon _{rs} }} = \frac{1}{{2\varepsilon (r + 1)}}\left[ {\frac{1}{{r - s}} + \frac{1}{{r + s + 2}}} \right]\,.
\end{equation}
Since $V$ is a real symmetric matrix, \eqref{equ.kzjql8y} is simply
\begin{equation}\label{equ.avhr7jt}
E_r^{(2)} = \sum_{s = 0}^{r - 1} {\frac{{V_{rs}^2 }}{{\varepsilon _{rs} }}}  + \sum_{s = r + 1}^\infty {\frac{{V_{rs}^2 }}{{\varepsilon _{rs} }}}\,.
\end{equation}
We note that the matrix elements $V_{rs}$ occuring in~\eqref{equ.avhr7jt} are necessarily \mbox{off-diagonal} (since $s\ne r$). Furthermore the only surviving elements $V_{rs}$, according to~\eqref{equ.jevxxpj}, are those for which $r$ and $s$ are both odd or both even. 

\bigskip

It therefore follows from~\eqref{equ.jevxxpj} that
\begin{equation}\label{equ.hnip2m6}
\begin{split}
V_{rs}  &={( - 1)^{(r + s)/2} c_1\left( {\frac{1}{{(r - s  )^2 }} - \frac{1}{{(r + s + 2)^2 }}} \right)} \\
&\qquad\qquad { - ( - 1)^{(r + s)/2} c_2\left( {\frac{1}{{(r - s )^4 }} - \frac{1}{{(r + s + 2)^4 }}} \right)}\,,
\end{split}
\end{equation}
where
\[c_1=\frac{{16\lambda a^4 }}{{\pi ^2 }}=\frac{\pi^2}{24}c_2\,.\]
Taking~\eqref{equ.cxdserf} into account, the summand in~\eqref{equ.avhr7jt} is therefore
\begin{equation}\label{equ.ifto5d5}
\begin{split}
\frac{{V_{rs} ^2 }}{{\varepsilon _{rs} }} &= \frac{1}{{2\varepsilon (r + 1)}}\left( {\frac{1}{{r - s}} + \frac{1}{{r + s + 2}}} \right)\\
&\quad \times \left( {\frac{1}{{(r - s)^2 }} - \frac{1}{{(r + s + 2)^2 }}} \right)^2\\ 
&\qquad \times \left( {c_1  - c_2 \left[ {\frac{1}{{(r - s)^2 }} + \frac{1}{{(r + s + 2)^2 }}} \right]} \right)^2\,. 
\end{split}
\end{equation}
The sum in~\eqref{equ.avhr7jt} is easier to evaluate if the energy eigenvalues are grouped by parity:
\[
E_{2r}^{(2)} = \sum_{s = 0}^{2r - 1} {\frac{{V_{2r,s}^2 }}{{\varepsilon _{2r,s} }}}  + \sum_{s = 2r + 1}^\infty {\frac{{V_{2r,s}^2 }}{{\varepsilon _{2r,s} }}}
\]
and
\[
E_{2r+1}^{(2)} = \sum_{s = 0}^{2r} {\frac{{V_{2r+1,s}^2 }}{{\varepsilon _{2r+1,s} }}}  + \sum_{s = 2r + 2}^\infty {\frac{{V_{2r+1,s}^2 }}{{\varepsilon _{2r+1,s} }}}\,,
\]
for quantum number $r=0,1,2,\ldots$

\bigskip

Using the summation identity (equation 2.6 of~\cite{gould})
\[
\sum_{k = q}^n {f_k }  = \sum_{k = \left\lfloor {(q + 1)/2} \right\rfloor }^{\left\lfloor {n/2} \right\rfloor } {f_{2k} }  + \sum_{k = \left\lfloor {(q + 2)/2} \right\rfloor }^{\left\lfloor {(n + 1)/2} \right\rfloor } {f_{2k-1} },\quad n\ge q+1\,, 
\]
where $\lfloor p\rfloor$ denotes the floor of $p$, that is, the greatest integer less than or equal to $p$, the above sums can be expressed as
\begin{equation}\label{equ.hxqejv7}
E_{2r}^{(2)}  = \sum_{s = 0}^{r - 1} {\frac{{V_{2r,2s}^2 }}{{\varepsilon _{2r,2s} }}}  + \sum_{s = r + 1}^\infty  {\frac{{V_{2r,2s}^2 }}{{\varepsilon _{2r,2s} }}} 
\end{equation}
and
\begin{equation}\label{equ.i2rivoe}
E_{2r + 1}^{(2)}  = \sum_{s = 0}^{r - 1} {\frac{{V_{2r + 1,2s + 1}^2 }}{{\varepsilon _{2r + 1,2s + 1} }}}  + \sum_{s = r + 1}^\infty  {\frac{{V_{2r + 1,2s + 1}^2 }}{{\varepsilon _{2r + 1,2s + 1} }}}\,. 
\end{equation}
Maple is able to evaluate the sums in~\eqref{equ.hxqejv7} and~\eqref{equ.i2rivoe}, with the appropriate summand in each case obtained from~\eqref{equ.ifto5d5}, and we have (see the Maple code in the appendix)
\[
\begin{split}
E_{2r}^{(2)}  &= \frac{{ma^{10} \lambda ^2 }}{{\hbar ^2 }}\left[ {\frac{{32}}{{225\pi ^2 (2r + 1)^2 }} - \frac{{1184}}{{35\pi ^4 (2r + 1)^4 }}} \right.\\
&\qquad\left. { + \frac{{10048}}{{5\pi ^6 (2r + 1)^6 }} - \frac{{44928}}{{\pi ^8 (2r + 1)^8 }} + \frac{{278784}}{{\pi ^{10} (2r + 1)^{10} }}} \right]
\end{split}
\]
and
\[
\begin{split}
E_{2r + 1}^{(2)}  &= \frac{{ma^{10} \lambda ^2 }}{{\hbar ^2 }}\left[ {\frac{8}{{225\pi ^2 (r + 1)^2 }} - \frac{{74}}{{35\pi ^4 (r + 1)^4 }}} \right.\\
&\qquad\left. { + \frac{{157}}{{5\pi ^6 (r + 1)^6 }} - \frac{{351}}{{2\pi ^8 (r + 1)^8 }} + \frac{{1089}}{{4\pi ^{10} (r + 1)^{10} }}} \right]\\
&\\
 &= \frac{{ma^{10} \lambda ^2 }}{{\hbar ^2 }}\left[ {\frac{{32}}{{225\pi ^2 (2r + 2)^2 }} - \frac{{1184}}{{35\pi ^4 (2r + 2)^4 }}} \right.\\
&\qquad\left. { + \frac{{10048}}{{5\pi ^6 (2r + 2)^6 }} - \frac{{44928}}{{\pi ^8 (2r + 2)^8 }} + \frac{{278784}}{{\pi ^{10} (2r + 2)^{10} }}} \right]\,,
\end{split}
\]
from which it follows that
\begin{equation}\label{equ.aabmsh0}
\begin{split}
E_r^{(2)}  &= \frac{{ma^{10} \lambda ^2 }}{{\hbar ^2 }}\left[ {\frac{{32}}{{225\pi ^2 (r + 1)^2 }} - \frac{{1184}}{{35\pi ^4 (r + 1)^4 }}} \right.\\
&\qquad\left. { + \frac{{10048}}{{5\pi ^6 (r + 1)^6 }} - \frac{{44928}}{{\pi ^8 (r + 1)^8 }} + \frac{{278784}}{{\pi ^{10} (r + 1)^{10} }}} \right]\,.
\end{split}
\end{equation}

Adding~\eqref{equ.zdjzs3r}, \eqref{equ.g606140} and \eqref{equ.aabmsh0}, we finally obtain
\[
\begin{split}
E_r  &\approx \frac{{\pi^2 \hbar^2 }}{{8ma^2 }}(r + 1)^2  + \frac{{\lambda a^4 }}{5}\left[ {1 - \frac{{20}}{{\pi^2 (r + 1)^2 }} + \frac{{120}}{{\pi^4 (r + 1)^4 }}} \right]\\
&\qquad + \frac{{32m\lambda ^2 a^{10} }}{{\hbar ^2 }}\left[ {\frac{1}{{225\pi^2 (r + 1)^2 }} - \frac{{37}}{{35\pi^4 (r + 1)^4 }}} \right.\\
&\qquad\qquad\qquad\qquad\left.{  + \frac{{314}}{{5\pi^6 (r + 1)^6 }}- \frac{{1404}}{{\pi^8 (r + 1)^8 }} + \frac{{8712}}{{\pi^{10} (r + 1)^{10} }}}\right]\,,
\end{split}
\]
as an approximate expression for the eigenvalues of the bounded quartic oscillator.

\section{Summary and conclusion}
Using the Rayleigh-Schr\"odinger perturbation theory and with the aid of a summation identity and the Computer Algebra System Maple, we have derived an approximate expression for the energy eigenvalues of the bounded quartic oscillator. This is the same expression that was obtained much earlier in reference~\cite{fernandez} through a more complicated approach. Similar results to ours are also contained in reference~\cite{navarro} where exact diagonalization was done and perturbative series up to the third order in $\lambda$ were also developed for the energy levels. However, the RS sums were determined numerically in that paper as, apparently, closed form could not be found for them; and furthermore only results for the ten lowest eigenvalues were computed.
\section*{Appendix}
\appendix
\setcounter{secnumdepth}{0}

\section{Maple code to evaluate $E_r^{(2)}$}

\begin{verbatim}
> V:=(r,s)->(-1)^(r+s)*c1*(1/(r-s)^2-1/(r+s+2)^2)
  -(-1)^(r+s)*c2*(1/(r-s)^4-1/(r+s+2)^4):
> epsilon:=(r,s)->epsilon*(r+s+2)*(r-s):
> simplify(expand(eval(V(r,s)^2/epsilon(r,s),[r=2*r,s=2*s])))
  assuming r,posint,s,posint,c1>0,c2>0:
> summand:=convert(%,parfrac,s):
> S1:=sum(summand,s=0..r-1) assuming r,posint,s,posint,c1>0,c2>0:
> S2:=sum(summand,s=r+1..infinity) assuming r,posint,s,posint,c1>0,c2>0:
> ssum:=simplify(S1+S2):
> fsum:=expand(eval(ssum,[epsilon=Pi^2*h^2/8/m/a^2,
  c1=16*lambda*a^4/Pi^2,c2=16*24*lambda*a^4/Pi^4])):
> arranged:=collect(fsum,Pi):
> terms:=[seq(factor(op(i,arranged)),i=1..nops(arranged))]:
\end{verbatim}
\verb|> add(L,L=terms);|
{\tiny
\[
 {\frac{{32ma^{10} \lambda ^2 }}{{225\pi ^2 (2r + 1)^2 \hbar ^2 }} - \frac{{1184ma^{10} \lambda ^2 }}{{35\pi ^4 (2r + 1)^4 \hbar ^2 }} + \frac{{10048ma^{10} \lambda ^2 }}{{5\pi ^6 (2r + 1)^6 \hbar ^2 }} - \frac{{44928ma^{10} \lambda ^2 }}{{\pi ^8 (2r + 1)^8 \hbar ^2 }} + \frac{{278784ma^{10} \lambda ^2 }}{{\pi ^{10} (2r + 1)^{10} \hbar ^2 }}} 
\]
}

\end{document}